\documentclass[aps,prc,superscriptaddress,showpacs,amsmath,floatfix]{revtex4}

\usepackage{color}
\usepackage{graphicx}
\usepackage{bm}
\newcommand{\ba}{\begin{array}}
\newcommand{\ea}{\end{array}}

\def\ra{\rangle}
\def\la{\langle}
\def\vr{\vec{r}}
\def\wlow{\raisebox{-0.6ex}{\tiny \rm W}}
\def\clow{\raisebox{-0.6ex}{\scriptsize \rm ch}}
\def\wlowF{\raisebox{-0.4ex}{\!\tiny \rm W}}
\def\clowF{\raisebox{-0.4ex}{\!\scriptsize \rm ch}}

\newcommand{\vnabla}{\vec{\nabla}}

\newcommand{\vz}{\vec{z}}
\newcommand{\vs}{\vec{s}}
\newcommand{\vx}{\vec{x}}
\newcommand{\vy}{\vec{y}}
\newcommand{\vq}{\vec{q}}
\def\ra{\rangle}
\def\la{\langle}
\def\vr{\vec{r}}

\def\haf{\textstyle{1\over2}}
\def\lsim{\mathrel{\raise3pt\hbox to 8pt{\raise -6pt\hbox{$\sim$}\hss{$<$}}}}

\def\ra{\rangle}
\def\la{\langle}

\def\haf{\textstyle{1\over2}}
\def\lsim{\mathrel{\raise3pt\hbox to 8pt{\raise -6pt\hbox{$\sim$}\hss{$<$}}}}
\def\rsim{\mathrel{\raise3pt\hbox to 8pt{\raise -6pt\hbox{$\sim$}\hss{$>$}}}}

\begin{document}

\title{Nuclear Zemach Moments and Finite-Size Corrections to Allowed Beta Decay}

\author{X.B. Wang}
\affiliation{School of Science, Huzhou University, Huzhou
313000, China}
\author{J.L. Friar}
\affiliation{Los Alamos National Laboratory, Los Alamos, NM, USA  87545}
\author{A.C. Hayes}
\affiliation{Los Alamos National Laboratory, Los Alamos, NM, USA  87545}
\begin{abstract}
The finite-size correction to $\beta$-decay plays an important role in determining the expected antineutrino spectra from reactors at a level that is important for the reactor-neutrino anomaly. Here we express the leading-order finite-size correction to allowed $\beta$-decay
 in terms of Zemach moments. We calculate the Zemach moments within a Hartree-Fock model using a Skyrme-like energy density functional. We find that the Zemach moments are increased relative 
to predictions based on the simple assumption of identical uniform nuclear-charge and weak-transition densities.
However, for allowed ground-state to ground-state transitions in medium and heavy nuclei, the detailed nuclear structure calculations do not change
the finite-size corrections significantly from the simple model predictions, and are only 10-15\% larger than the latter even though the densities differ significantly.
\end{abstract}


\date{\today}
\maketitle

\section{Introduction}
The finite-size (FS) of the nucleus results in small corrections to several low-energy electromagnetic and weak interaction
processes,  such as precision studies of atomic hyperfine splitting\cite{zemach} and the Lamb shift\cite{friar1}, as well as nuclear beta decay\cite{hayes1}.
In all of these processes Zemach\cite{zemach} moments enter, and the magnitudes of some of these Zemach moments have been extracted \cite{Sick,extract}
from experimental measurements in very light nuclei. 
For heavier nuclei  finite-size effects are often parameterized with a simple estimate based on the size of a hypothetical uniform-density nucleus (viz., $R = r_0A^{1/3}$), 
where $r_0$ is typically taken to be 1.2 fm and $A$ is the nucleon number.
The FS correction to beta decay plays a significant role in 
the current reactor neutrino anomaly \cite{anomaly} and so it is important to examine the approximations and uncertainties for these corrections.
In the present work, we express the FS correction to allowed beta decay in terms of Zemach moments and 
compare nuclear structure calculations for these moments across a broad range of  masses with the often used $R=r_0A^{1/3}$  uniform-density approximation.
We note that forbidden $\beta$-decay transitions represent \cite{hayes1, hayes2, hayes3, bnl, Fallot} about 30\% of reactor antineutrino spectra and the FS corrections for these transitions are more complicated and more poorly understood than for allowed transitions.  We do not discussed Zemach moments for forbidden $\beta$-decays in the present work.

\label{sec2}
\section{The Finite-Size Correction to  Allowed Beta-Decay}
The  interaction of the out-going electron with the charge of the daughter nucleus is a large correction that 
has to be taken into account in  nuclear $\beta$-decay studies.
The primary correction involves replacing the plane wave solution
for the out-going electron with a point-Coulomb wave function. Standard practice involves
 introducing the Fermi function $F(E,Z,A)$ for a point-like nuclear charge, and then improving on this through a finite-size correction to $F$. Here $E$ is the total energy of the out-going electron with mass $m$, and $Z$ is the charge of the daughter nucleus. 
In the present work, we focus on the FS corrections to  allowed $\beta$-decay. 
This correction has been derived previously by other authors \cite{bible, holstein} in various approximations, but the focus of the present work is to formulate the problem in terms of the Zemach moments that
play an important role in other electromagnetic and weak interaction processes.
  
The electron density near the nuclear surface is increased by the attractive Coulomb interaction, so that the $\beta$-decay rate
is increased by the point-like Fermi correction.
The finite-size correction to the Fermi function decreases the electron density near the nucleus and the rate is then decreased (relative to
the point-nucleus Fermi function). Thus, if we express the FS correction in the form
$F\rightarrow F(1+\delta_{FS})$, the sign of the finite-size correction $\delta_{FS}$ is always negative. 
In the case of reactor fission beta-decay spectra, where the beta spectrum of each fission fragment is normalized to unity, the FS correction results in a decrease (increase) in the magnitude 
of the high-energy component of the electron (antineutrino) spectra. This increase in the high-energy component of
the antineutrino spectrum is a major source
 of the reactor neutrino anomaly.

The dimensionless parameters in the FS problem are $Z \alpha$ (where  $\alpha$ is the fine-structure constant),  ${\cal{E}} R/\hbar c $, where ${\cal{E}}$ is any of the energy scales in the problem (such as the electron mass, $m$, or energy, $E$, or the antineutrino energy, $E_\nu$), and $R$ is a generic nuclear size. The quantity $Z \alpha$ is always less than 1, ${\cal{E}} R/\hbar c \lsim 1/10$ for typical values of ${\cal{E}} \lsim 5$ MeV and $R \sim 4$ fm. We will restrict ourselves to first order in both $Z \alpha$ and the nuclear size $R$. 

The finite-size corrections to this order for allowed Fermi and Gamow-Teller decays were previously derived by Holstein\cite{holstein}. Two normalized finite-size densities contribute to the correction. One is the weak transition density $\rho\wlow (r)$ that specifies the probability of finding  the beta-decaying nucleon at $r$. In a simple shell model this density is the product 
of the initial and final radial single-particle wave functions of the decaying nucleon. 
The second density is the charge density of the nucleus, $\rho\clow (r)$, which determines the Coulomb interaction. To the order that we work there is one density of each type.

The quantity $\Delta h_1$ in Eqn.~(25) of Holstein's paper is equivalent to our $\delta_{FS}$. His form-factor integrals (labeled $X$ and $Y$ in that equation) can be converted to our Zemach moments using $X+Y = -\frac{\pi}{8}\la r \ra_{(2)}$ and $X= -\frac{\pi}{8}\la r \ra_{(2)}^r$.
Equation~(2) of Ref.~\cite{friar2} expresses  $\la r \ra_{(2)}$ in terms of weak and charge form factors: $\la r \ra_{(2)} = -\frac{4}{\pi}\int_0^\infty \frac{dq}{q^2} (F\wlowF (q^2) F\clowF (q^2) - 1)$, and $\la r \ra_{(2)}^r$ is similar (see the Appendix). Our Zemach moments are given by
\begin{equation}
\la r \ra_{(2)}  = \int d^3r \, \rho\wlow (r) \int d^3 s \, \rho\clow (s) \mid \vr -\vs \mid = \int d^3s \; s  \int d^3 r\; \rho\wlow (r)\, \rho\clow (\mid \vr -\vs \mid)\,  ,\label{Z1}
\end{equation}
which is the first radial moment of the convoluted or Zemach density $\rho_{(2)} (s)$, and 
\begin{equation}
 \la r \ra_{(2)}^r  = \int d^3r \, \rho\wlow (r) \;\vr \cdot \vnabla_r \int d^3 s \, \rho\clow (s) \mid \vr -\vs \mid = \int d^3 s \; s \int d^3 r\,  \rho\wlow (r)\; r \,\frac{\partial}{\partial r} \rho\clow (\mid \vr -\vs \mid)\, ,\label{Zr}
\end{equation}
which is the first moment of a more complicated Zemach-type density $\rho_{(2)}^r (s)$ (see the Appendix). Inserting the above translations of $X$ and $Y$ in terms of  Zemach moments into Holstein's Eqn.~(25) we find that the fractional FS correction to allowed Gamow-Teller transitions of the desired order is
\begin{equation}
\delta_{FS}= -\frac{\, Z\, \alpha}{3\hbar c}\,\left ( 4E\, \la r \ra_{(2)} + E\, \la r \ra_{(2)}^r -\frac{E_\nu \la r \ra_{(2)}^r }{3} + \frac{m^2c^4}{E} (2\, \la r \ra_{(2)} - \la r \ra_{(2)}^r) \right ) \; . \label{FSZ}
\end{equation}
We have independently calculated $\delta_{FS}$ using different techniques than Ref.~\cite{holstein} and have verified Eqn.~(\ref{FSZ}), which is our full result for the  finite-size correction to allowed  Gamow-Teller $\beta$-decay  expressed in terms of
 two Zemach moments, $\la r \ra_{(2)}$ and $\la r \ra_{(2)}^r$,  both of which are nuclear-structure dependent. 

In the special case where $ \rho\wlow$ and $ \rho\clow$ are identical, one can show that 
$\la r \ra_{(2)}^r\longrightarrow \frac{1}{2} \la r \ra_{(2)} $.
 In that very useful limit we have
\begin{equation}
\delta_{FS}= -\frac{3}{2} \frac{Z\, \alpha}{\hbar c} \, \la r \ra_{(2)}\,\left ( E -\frac{E_\nu}{27} + \frac{m^2 c^4}{3 E}  \right ) \;  \label{deltafs}
\end{equation}
for allowed Gamow-Teller transitions. 
This is the form  of the FS correction that was used in ref.~\cite{hayes1}, where the additional assumption
of   a uniform distribution of radius $R$ for both  the weak transition and
charge densities was made. Under this assumption\cite{friar1},  $\left<r\right>_{(2)}=\frac{36}{35}R$, where $R \cong1.2\;A^{1/3}$fm is the radius of a uniform-density charge distribution that is obtained by fitting the root-mean-square charge radii of nuclei across the periodic table\cite{nix}.

For allowed Fermi transitions (as opposed to allowed Gamow-Teller) 
one should replace the numerical coefficient of $E_\nu$ by +1/9 in Eqn.~(\ref{deltafs}) 
and by +1 in Eqn.~(\ref{FSZ}). 
Both results agree with the first of Eqns.~(25) in Ref.~\cite{holstein}, where $| a |^2$ and $| c |^2$ denote the squares of the Fermi and Gamow-Teller matrix elements, respectively, and play no role in extracting the Coulomb corrections.
This also agrees with Bottino and Ciocchetti\cite{BC} for allowed Gamow-Teller transitions involving a common monopole form factor (i.e., a Yukawa  distribution), 
which was the only distribution they treated in detail. 
The allowed Fermi $\beta^+$-transitions considered in Ref.~\cite{DN} also agree with Eqn.~(\ref{FSZ})   
for the modified Gaussian distribution considered there, if we take account of the sign change needed to treat $\beta^+$-decay rather than $\beta^-$-decay.
Reference~\cite{AK} used a point-like charge distribution and a uniform nuclear transition density, 
which corresponds to $ \la r \ra_{(2)}^r =  \la r \ra_{(2)} \longrightarrow  \la r \ra = \frac{3\, R}{4}$ 
and agrees with Eqn.~(\ref{FSZ}) if one uses the conversion factors in Ref.~\cite{holstein}. 
We note that Eqn.~(\ref{deltafs}) differs from the corresponding one in Ref.~\cite{mueller}.

Most calculations of the FS correction appearing in the literature assume that the weak transition and charge densities are the same. 
However, the charge densities are determined by summing the distributions of all protons in the daughter nuclei, while the
 weak transition densities are determined by the product of the
 neutron and proton wave functions involved in the beta transitions. 
Thus we need to examine the effect of including more realistic
nuclear structure descriptions for these  density distributions in evaluating the Zemach moments appearing in Eqn.~(\ref{FSZ}).
For this purpose we use charge and weak transition densities derived from a density-functional-theory description of the nucleus. 

Although the integrals in Eqns.~(\ref{Z1}) and (\ref{Zr}) appear very complicated, in reality each is no worse than a triple integral. Moreover, each can be expressed as the simple $r$-moment of densities $\rho_{(2)} (r)$ and $\rho_{(2)}^r (r)$, respectively. We demonstrate this in the Appendix: Zemach Moments and Densities.

\section{Nuclear density functional theory}

Density functional theory (DFT) is based on theorems that prove the existence of energy functionals 
for many-body systems, which, in principle, include all many-body correlations~\cite{Hoh1964,Lev1979,Koh1965}. 
Nuclear DFT has the advantage of being applicable across the entire chart of the nuclides. 
The main ingredient in nuclear DFT is the energy density functional 
that depends on the distributions of the nucleonic matter, spin, and kinetic densities, 
as well as their gradients \cite{Ben2003rmp}.  
The present work is based on the Skyrme-like energy density functional, which has the form
\begin{equation}\label{sky}
\mathcal{E}^{\text{Skyrme}} = \sum_{t=0,1} \int d^3 r \left(
\mathcal{H}_t^{\text{even}}({\boldsymbol r}) +
\mathcal{H}_t^{\text{odd}}({\boldsymbol r}) \right) \, ,
\end{equation}
where the time-even part of the energy density is defined in Ref.~\cite{Eng1975}, and the present calculations
do not include the time-odd component.

In the Hartree-Fock (HF) approximation
the single-particle states are calculated by solving the eigen-equations
\begin{eqnarray} \label{hte2}
\it h_{\alpha} \varphi_{i,\alpha}= \it e_{i,\alpha}  \varphi_{i,\alpha}\, ,
\end{eqnarray}
where $i$ indicates all nucleonic quantum numbers, $\alpha$ is used to distinguish the neutron ($ n$) or the proton ($ p$), and $\it e_{i,\alpha}$ is the corresponding eigen-energy.
If we assume spherical symmetry\cite{Dob1984}, the single-particle wave functions (SPWF) $\varphi_{i,\alpha}$  have good quantum numbers  $[n\ell j]$.
The (reduced) radial wave function of the $[n\ell j]$ orbital $u_{\alpha}(n\ell j,r)$ is normalized according to $\int dr\, u^2_{\alpha}(n\ell j,r) = 1$.
The local neutron and proton densities are then simply obtained from the radial wave functions as 
\begin{eqnarray}
\rho_{\alpha}(r)&=&\displaystyle
 \sum_i\la\Psi\mid a^\dag_{r,i\alpha}a_{r,i\alpha}\mid\Psi\ra\\
 &=&
  \frac{1}{4\pi r^2}\sum_{n\ell j}(2j+1)u_{\alpha}^2(n\ell j,r)\,,\hfill
\end{eqnarray}
where $\Psi$ is the total wave function, $i$ indicates all of the single-particle quantum numbers, and $\alpha$  distinguishes between neutrons and protons.

Many of the fission fragments dominating the reactor antineutrino spectra are deformed nuclei, and might not  
be well described by our Hartree-Fock model. 
However, our goal  for these nuclei is to gain an understanding of how FS corrections are affected by using nuclear structure models
that are considerably more sophisticated
than the assumption of equal uniform models for the weak transition density and the charge density.
\subsection{Charge and weak transition density}
In order to study the Zemach moments we require normalized charge and weak transition densities.
The normalized charge density is simply
\begin{eqnarray}\label{shf-charge}
\rho\clow(r)&=&\displaystyle
 \frac{ \rho_{p}(r)}{ \int d^3 r \,\rho_{p}(r)}
 =\frac{1}{Z} \rho_{p}(r)\, .\hfill
\end{eqnarray}
Within the Skyrme-Hartree-Fock model, the weak transition density simply involves one term and is given by
\begin{equation}
\rho\wlow(r) = \la\Psi_f\mid a^\dag_{r,kp}a_{r,ln}\mid\Psi_i\ra\, ,
\end{equation}
where $\Psi_i$ and $\Psi_f$ are the total wave functions for the parent and daughter nuclei, respectively, and $a_{r,ln}$ is the  annihilation operator 
for the single neutron with quantum numbers $l$ in the parent nucleus and $a^\dag_{r,kp}$ is the creation operator for the single proton with quantum numbers $k$ in the daughter nucleus. 
For allowed $\beta$-decay transitions, the neutron or proton are often in the same $[n\ell j]$ orbital or in   
spin-orbit-partner orbitals (i.e., orbitals with the same $n\ell$ and $j \,, j' =\ell \pm 1/2 $). 
The weak transition density is then given by the product of reduced wave functions
 \begin{eqnarray}\label{shf-weak}
\rho^{\rm SHF}_{\rm W}(r)&=&\displaystyle
\frac{1}{4\pi r^2} \frac{u_{p}^*(n\ell j',r) u_{n}(n\ell j,r)}{ \int  dr\, u_{p}^*(n\ell j',r) u_{n}(n\ell j,r)}. \hfill
\end{eqnarray}
We note that, since the spherical mean fields derived from the Skyrme density functional involve both a spin-orbit  and Coulomb interaction,
  the neutron and proton radial wave functions are different even for states in the same $[n\ell j]$ orbital, but are not orthogonal.
\section{Results and discussion}
\label{sec3}
 For the purpose of the present study,  the numerical code {\sc HFBRAD}~\cite{hfbrad} was used to solve the Skyrme-Hartree-Fock equations in coordinate space within the assumption of spherical symmetry. The energy density functional SLy4 Skyrme-EDF~\cite{Cha1998} was used, which is a common choice for DFT predictions. 
\subsection{Density distribution}
 \begin{figure}
\includegraphics[angle=0,width=12 cm]{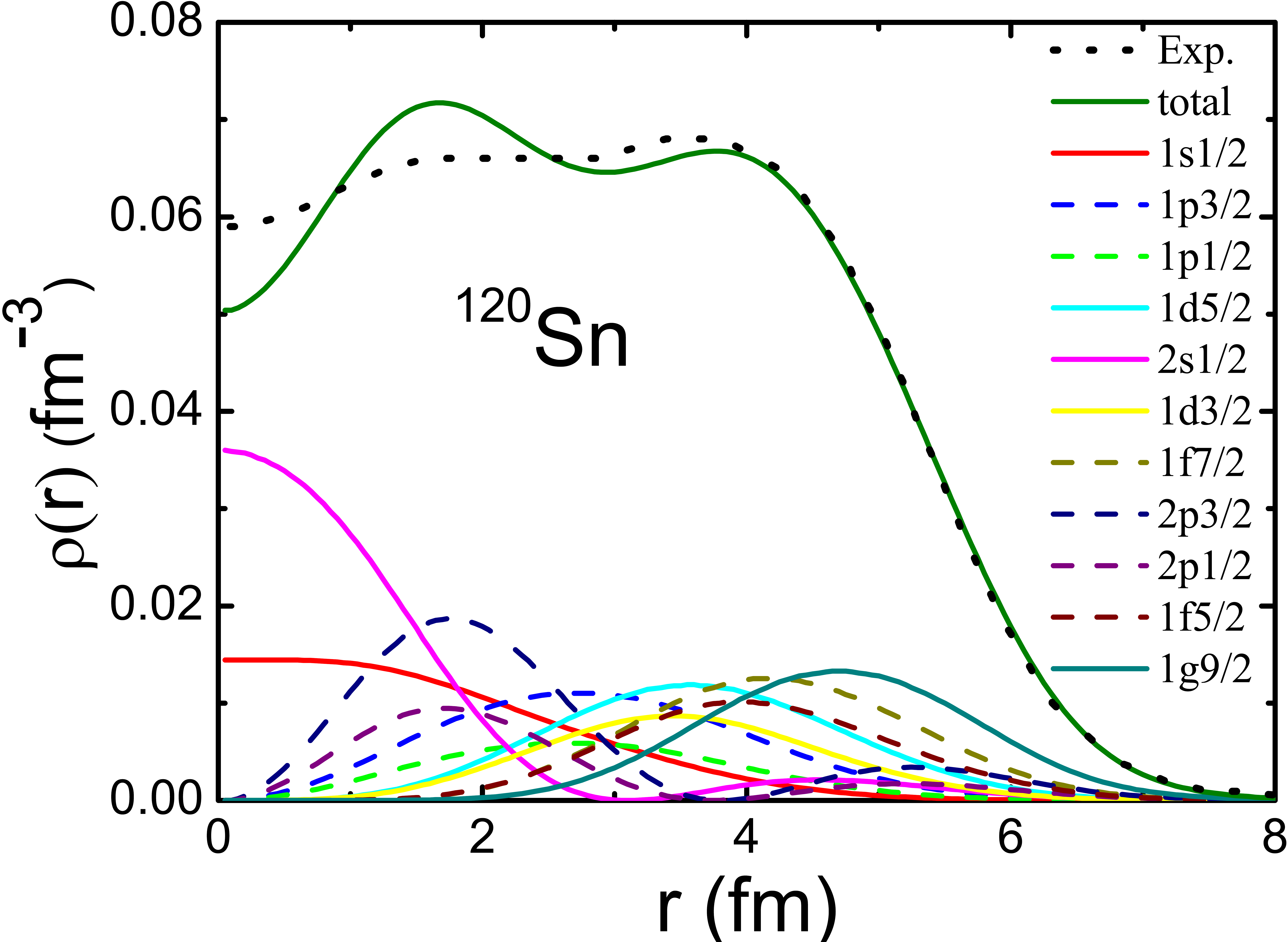}
\caption[T]{\label{fig:charge-density-sn120}
(Color online) Charge density distribution of $^{120}$Sn as calculated by SLy4 EDF. The contribution from each single-particle state is given. The experimental data~\cite{STer08} are also plotted for comparison.}
\end{figure}
 
 In Fig.~(\ref{fig:charge-density-sn120})
we show the SLy4 Skyrme-EDF prediction for the total charge density of $^{120}$Sn, 
as well as the contributions from individual shells.
 The theoretical charge density distribution is in reasonable agreement with  the experimental one, especially in the nuclear
surface. 
Orbitals with higher angular momentum tend to peak near the nuclear surface and thus dominate this region of the charge density. 
At smaller (intermediate) radii  the experimental charge density is relatively flat. In contrast,
the intermediate theoretical density exhibits features that can largely  be attributed to orbitals with nodes, which tend to peak
at these radii.
This effect has been discussed in detail in refs.~\cite{negele}, where it is found that Hartree-Fock-type models tend to  predict
exaggerated charge density fluctuations, in part because of an inadequate treatment of many-body correlations.

Except for $s$-wave-to-$s$-wave transitions, the weak transition densities are zero at the origin, and they
 peak at intermediate radii. Thus, they are distinctly different in shape from the charge densities, 
in strong contrast to the uniform density assumption. 
\subsection{Charge, weak, and convoluted densities}
\begin{figure}
\includegraphics[angle=0,width=12 cm]{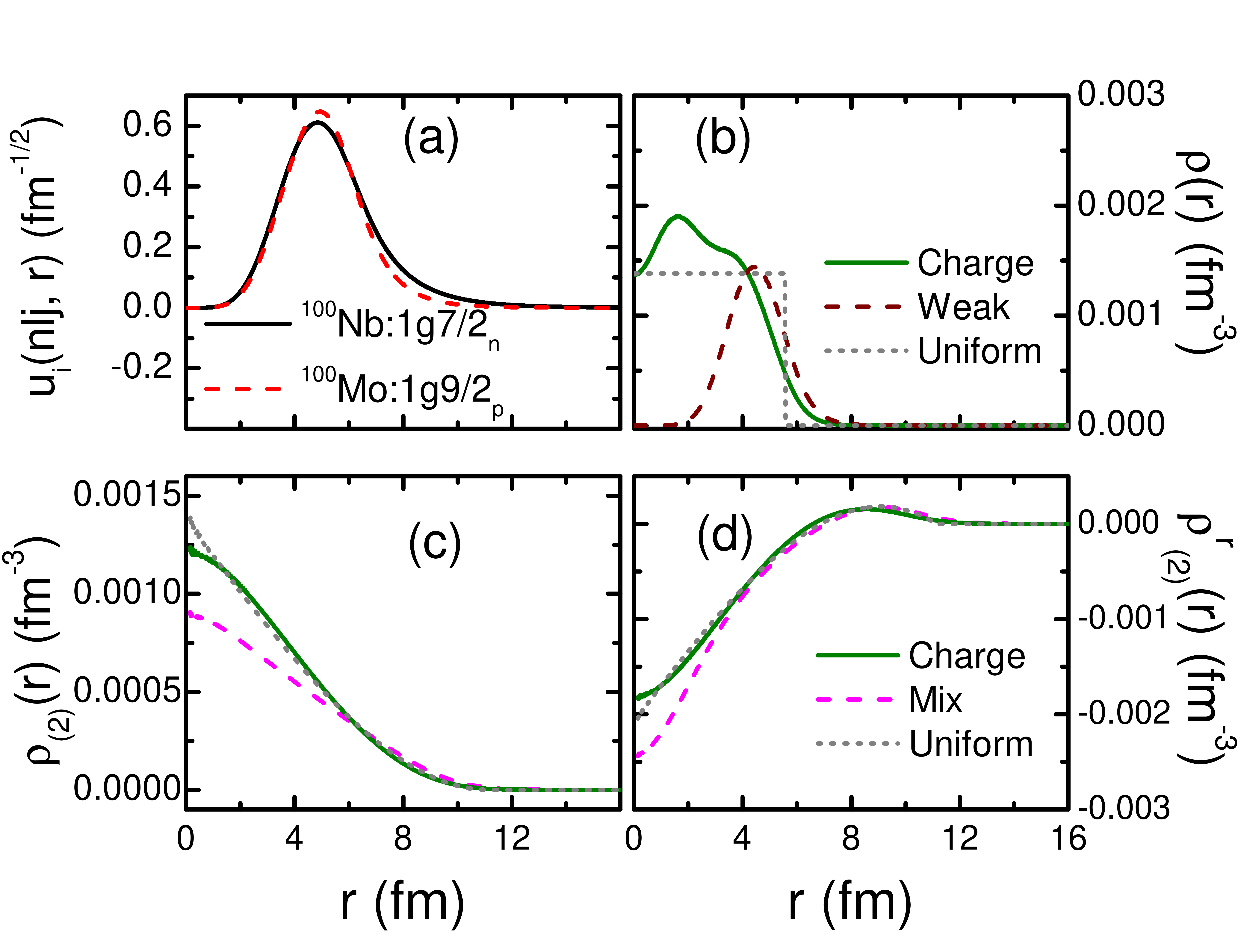}
\caption[T]{\label{fig:density-Nb100}
(Color online) (a): The radial wave functions of the neutron in the parent nucleus and the proton in the daughter nucleus participating in the allowed ground-state-to-ground-state $\beta^-$-decay of  $^{100}$Nb; (b): Normalized charge density in the daughter nucleus $^{100}$Mo, normalized weak transition density calculated from the corresponding SPWFs, and the uniform density distribution for the A=100 nucleus,  labeled by ``Charge", ``Weak," and ``Uniform," respectively. The convoluted densities $\rho_{(2)}(r)$ and $\rho^r_{(2)}(r)$ determining the Zemach moments $ \la r \ra_{(2)} $ and $ \la r \ra^r_{(2)} $, respectively,  are plotted in (c) and (d). Assuming that the weak density is identical to the normalized charge density produces the curve labeled ``Charge."  The curve labeled ``Mix" uses the calculated charge and weak densities, while  the curve labeled ``Uniform" assumes that both densities have a common uniform density distribution.}
\end{figure}

\begin{figure}
\includegraphics[angle=0,width=12 cm]{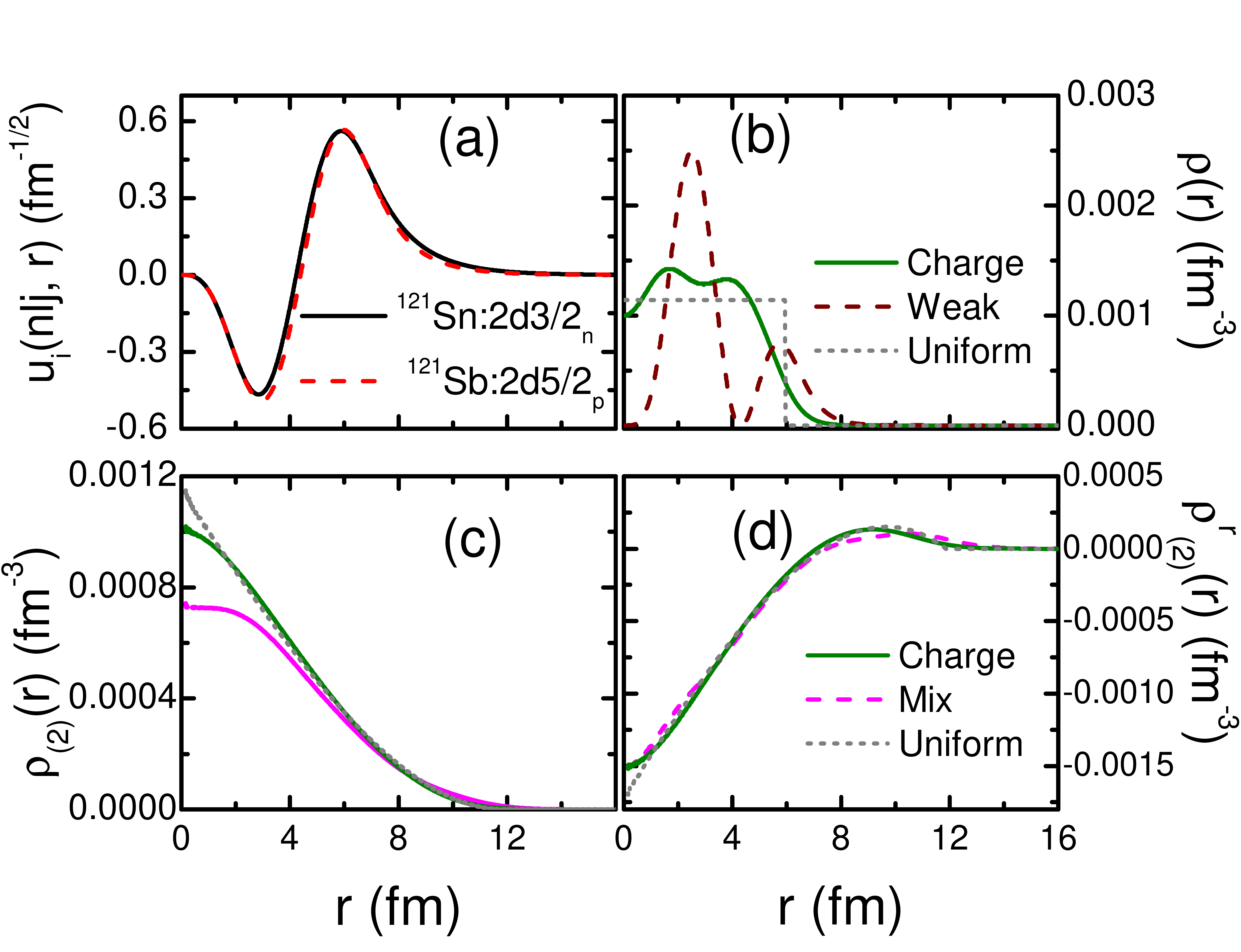}
\caption[T]{\label{fig:density-Sn121}
(Color online) The same as Fig.~(\ref{fig:density-Nb100}), but for the allowed ground-state-to-ground-state $\beta^-$-decay of  $^{121}$Sn.}
\end{figure}

In Figs.~(\ref{fig:density-Nb100}) and (\ref{fig:density-Sn121})
we compare the charge and weak transition densities for the parent/daughter pairs $^{100}$Nb/$^{100}$Mo and
$^{121}$Sn/$^{121}$Sb, respectively, where the transition densities represent 
the allowed ground-state-to-ground-state $\beta$-decays. 
For both nuclear systems the charge and weak transition densities are very different in shape.
Panel (a) shows the single-particle wave functions of the last neutron and the last proton that are
closest to the Fermi surfaces.
The corresponding normalized weak transition density, calculated using Eqn.~(\ref{shf-weak}), is
shown in panel (b), and is compared with the normalized charge density distribution and the uniform density distribution. 

The densities that determine
the Zemach moments $\la r \ra_{(2)}$ and $\la r \ra_{(2)}^r$ are shown in panels (c) and (d), respectively. They are derived in the Appendix in Eqns.~(\ref{rho2x}) and (\ref{rho2rx}) in a form convenient for computation
\begin{equation}
\rho_{(2)} (r) =  \frac{4 \pi}{2 r} \int_0^{\infty} dx\, x \;  \rho\wlow (x) \,\int_{\mid x-r \mid}^{x+r} dz\; z\; \rho_ {\rm ch} (z)\, ,
\end{equation}
and
\begin{equation}
\rho_{(2)}^r (r)\, = -\rho_{(2)} (r) + \frac{1}{2r} \int d^3x \; \rho\wlow (x) \left[ (x+r) \rho_ {\rm ch} (x+r) -\;(x-r) \rho_ {\rm ch} (\mid x-r \mid) \right]\,. 
\end{equation}
We note that $\int d^3r\, \rho_{(2)} (r) = 1$ and $ \int d^3r \; \rho_{(2)}^r (r)\,= 0$, 
and that $ \la r^2 \ra_{(2)} = \la r^2 \ra\wlow + \la r^2 \ra\clow\;$,  where the first quantity is the mean-square radius of the weak density and the second quantity is the mean-square radius of the charge density.

The Skyrme-Hartree-Fock convoluted densities $\rho_{(2)} (r)$ and $\rho_{(2)}^r (r)$ 
are found not to deviate significantly from the uniform density predictions, contrary to naive expectations based on the large differences between the individual component densities.
The largest differences are seen at small radii, where the Skyrme-Hartree-Fock
 convoluted densities tend to be lower than those
obtained assuming uniform densities. The reason for this lies in the convolution procedure, which folds the charge and weak densities together. This  leads to greater Zemach-density components at large radii, as seen in the panels. The normalization condition weights the tails of densities more heavily, and this extra strength in the tails must be compensated by reducing strength near the origin. We note that folding together two uniform distributions (each with extent $R$) leads to a Zemach density that extends to $2R$ \cite{friar1}.
\subsection{Zemach moments}
In Table~I we list the moments $\la r\ra$, $\la r^2\ra$ and $\la\frac{1}{r} \ra$ derived from the Skyrme-Hartree-Fock point densities and compare 
these
to the predictions of a simple uniform density that extends to  $R=1.2A^{1/3}$.
The corresponding Zemach moments determining the FS correction are listed in Table~II.
The Hartree-Fock Zemach moments are larger in general than the uniform-density values.
But the value of $\la r\ra_{(2)}$ in heavy nuclei is only 10\% higher that the uniform-density estimate. The Hartree-Fock prediction for the
smaller Zemach-like moment $\la r \ra_{(2)}^r$ is  30\% higher than the uniform-density value, 
but $\la r \ra_{(2)}^r$ is significantly less important numerically in Eqn.~(\ref{FSZ}) than $\la r\ra_{(2)}$. 
The ratio $\la r\ra_{(2)}$/$\la r\ra_{(2)}^r$ for the mixed case is closer to 1.7 than the equal-density prediction of 2.0.

All of the nuclear structure information required to calculate the leading-order finite-size correction to $\beta$-decay is contained in the two Zemach moments, $ \la r \ra_{(2)}$ and $ \la r \ra_{(2)}^r$.
Our primary purpose in this work is to assess the quality of approximating the physical charge and weak densities by equal uniform densities, 
which is the approximation for the Zemach moments that was used in the calculations of Ref.~\cite{hayes1}. 
We can see from Figs.~(\ref{fig:density-Nb100}) and (\ref{fig:density-Sn121}) that the weak transition density looks nothing like the charge or uniform densities for those cases. Although the charge density has roughly the shape of a uniform density, the weak density has substantial components that peak in the tail of the charge distribution, and  Table~I demonstrates that the mean-square radius of the weak density is substantially greater than that of the charge density. 
If the mean-square radius of the weak density is its dominant feature, 
 the weak form factor can be approximated by a two-term Taylor series (viz., $F\wlowF (q^2) \cong 1-q^2 \la r^2 \ra\wlow /6$),
 which is an approximation used in Ref.~\cite{bible}. 
In the Appendix we show that this leads to an expression relating the Zemach and charge density moments, namely,
 $\la r \ra_{(2)}^{BB} \cong \la r \ra\clow + \la r^2 \ra\wlow  \la \frac{1}{r} \ra\clow/3$. The latter
 approximate  Zemach moment (labeled ``Mix BB" in Table~II) is greater than the Hartree-Fock value labeled ``Mix" by roughly 5\%, and is therefore fairly accurate. 
The Hartree-Fock ``Mix" value is larger that the equal uniform density result by amounts that 
varies from 10\% to 30\% from heavy to light nuclei, as stated above. 
Because the $\beta$-decaying nuclei in reactors tend to be rather heavy, the equal uniform density approximation is actually rather good, and the uncertainties ($\sim$ 10-15\%) in the values of $ \la r \ra_{(2)}$ are in general much less than those assumed in Ref.\cite{hayes1}. It is nevertheless remarkable that two densities with very different shapes produce Zemach moments in substantial agreement with the simple equal and uniform model.

\begin{table*}
\begin{tabular}{lcccccccccccccc }
\hline\hline
        &~~$ \la r \ra^{\rm Uni}$ ~~  &~~$ \la r \ra^{\rm Ch}$~~ &~~ $ \la r \ra^{\rm Weak} $ ~~& ~~$ \la r^2 \ra^{\rm Uni} $ ~~ &~~$ \la r^2 \ra^{\rm Ch} $ ~~&~~ $ \la r^2 \ra^{\rm Weak} $ ~~&~~ $ \la \frac{1}{\raisebox{0.2ex} r} \ra^{\rm Uni} $~~ & ~~$ \la \frac{1}{\raisebox{0.2ex} r} \ra^{\rm Ch} $ ~~&  \  \\   \hline
        $^{14}$C/$^{14}$N & 2.17 & 2.43 & 2.80 & 5.02 & 6.73 & 8.78 & 0.52 & 0.50 \\ 
        $^{25}$Na/$^{25}$Mg & 2.63& 2.79& 3.29 & 7.39 & 8.75 & 11.63 & 0.43 & 0.42 \\ 
        $^{35}$S/$^{35}$Cl & 2.94& 3.08& 3.57 & 9.24 & 10.66 & 13.77 & 0.38 & 0.39 \\ 
        $^{45}$Ca/$^{45}$Sc & 3.20& 3.30 & 4.03 & 10.93 & 12.06& 17.12 & 0.35 & 0.36 \\ 
        $^{61}$Cr/$^{61}$Mn & 3.54& 3.53 & 4.23& 13.39  & 13.74 & 18.88 & 0.32 & 0.33 \\ 
        $^{64}$Co/$^{64}$Ni & 3.60 & 3.61 & 4.23 & 13.82 & 14.29 & 18.87 & 0.31 & 0.32 \\ 
        $^{100}$Nb/$^{100}$Mo & 4.18 & 4.13 & 4.97 & 18.61& 18.74 & 25.80 & 0.27 & 0.28 \\ 
        $^{104}$Nb/$^{104}$Mo& 4.23 & 4.18 & 5.01 & 19.11 & 19.15 & 26.14& 0.27 & 0.28 \\ 
        $^{121}$Sn/$^{121}$Sb & 4.45& 4.42 & 4.98 & 21.14 & 21.33 & 28.11 & 0.25 & 0.26 \\ \hline\hline
\end{tabular}\caption{Single-density moments  $\la r \ra$,  $\la r^2 \ra$, and  $\la \frac{1}{ r} \ra$, with units fm, fm$^2$ and fm$^{-1}$, respectively. These moments are derived using a uniform-density distribution, a point charge density, or a weak transition density, which are labeled as ``Uni", ``Ch", and ``Weak", respectively.}
\end{table*}

\begin{table*}
\begin{tabular}{lccccccccccccccc }
\hline\hline
        &$ \la r \ra_{(2)}^{\rm Uni}$~  & $~ \la r \ra^{{\rm Uni}BB}_{(2)} $~ &~$\la r \ra_{(2)}^{\rm Ch}$ ~&~ $ \la r \ra^{\rm Mix}_{(2)} $ ~&~ $ \la r \ra^{{\rm Mix}BB}_{(2)} $~  &~$ \la r \ra^{r~\rm Uni}_{(2)} $ ~&~$ \la r \ra^{r~{\rm Uni}BB}_{(2)} $ ~&~ $ \la r \ra^{r~\rm Ch}_{(2)} $ ~&~ $ \la r \ra^{r~\rm Mix}_{(2)} $~ &~$ \la r \ra^{r~{\rm Mix}BB}_{(2)} $ &  \  \\   \hline
        $^{14}$C/$^{14}$N & 2.97&3.04& 3.40 & 3.66 & 3.89 & 1.49 &1.74& 1.69 & 2.09 &2.93  \\ 
        $^{25}$Na/$^{25}$Mg & 3.61&3.68& 3.89 & 4.22 & 4.44 & 1.80 &2.11& 1.94 & 2.47 &3.29  \\ 
        $^{35}$S/$^{35}$Cl & 4.04&4.12& 4.30& 4.62 & 4.86 & 2.02 & 2.36&2.14 & 2.66 & 3.57 \\ 
        $^{45}$Ca/$^{45}$Sc & 4.39&4.48& 4.58 & 5.07 & 5.33 & 2.20&2.56& 2.28 & 3.07 &4.07  \\ 
        $^{61}$Cr/$^{61}$Mn & 4.86&4.96& 4.90 & 5.36& 5.62  & 2.43 &2.83& 2.44 & 3.19 &4.17  \\ 
        $^{64}$Co/$^{64}$Ni & 4.94 &5.04& 5.00 & 5.41 & 5.64 & 2.47&2.88 &2.49 & 3.16 & 4.06 \\ 
        $^{100}$Nb/$^{100}$Mo & 5.73&5.85 & 5.73 & 6.28 & 6.55 & 2.86 &3.34& 2.85 & 3.76 &4.83  \\ 
        $^{104}$Nb/$^{104}$Mo& 5.80 &5.93& 5.79 & 6.33 & 6.60 & 2.90 &3.39 &2.89& 3.78 &  4.83\\ 
        $^{121}$Sn/$^{121}$Sb & 6.10&6.23& 6.11 & 6.56 & 6.86 & 3.05 &3.56&3.05 & 3.71& 4.87 \\ \hline\hline
\end{tabular}\caption{Zemach moments $ \la r \ra_{(2)} $ and $ \la r \ra^r_{(2)} $ calculated from convoluted densities $\rho_{(2)} (s)$ and  $\rho_{(2)} ^r(s)$, respectively, for the allowed ground-state-to-ground-state $\beta^-$-decays. The calculations using the uniform-density distribution, the charge density only, or charge and weak densities together, are labeled ``Uni", ``Ch", and ``Mix",  respectively. The values of $ \la r \ra^{\rm Uni}_{(2)} $ and $ \la r \ra^{\rm Mix}_{(2)} $ approximated by using Eqn.~(\ref{FSBB}) are labeled as $ \la r \ra^{{\rm Uni}BB}_{(2)} $and $ \la r \ra^{{\rm Mix}BB}_{(2)} $, respectively. The values of $ \la r \ra^{r~\rm Uni}_{(2)} $ and $ \la r \ra^{r~\rm Mix}_{(2)} $ approximated by using Eqn.~(\ref{FSXX}) are labeled as $ \la r \ra^{{r~\rm Uni}BB}_{(2)} $and $ \la r \ra^{{r~\rm Mix}BB}_{(2)} $, respectively.  All results are in units of fm.}
\end{table*}

\subsection{The finite-size corrections}
\begin{figure}
\includegraphics[angle=0,width=12 cm]{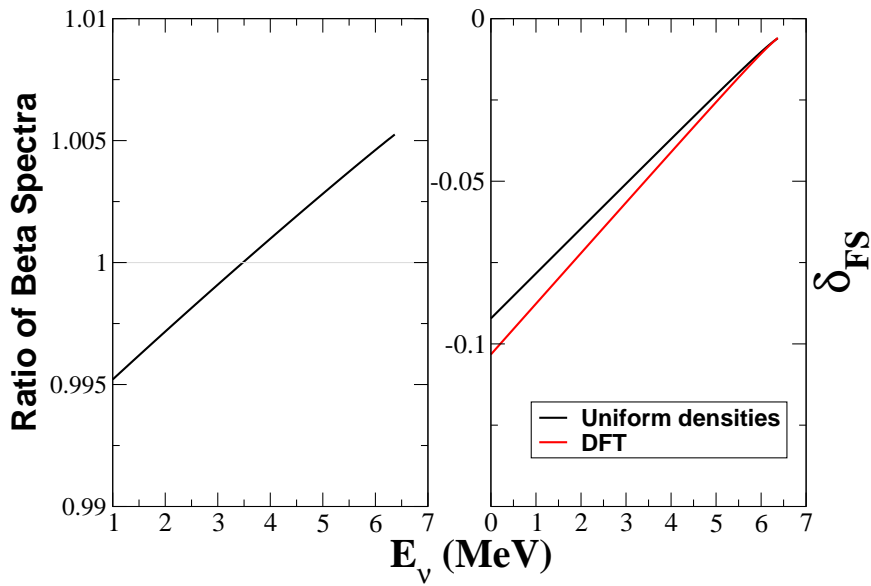}
\caption[T]{\label{fig:FS-100Nb}
(Color online) (Left) The ratio (DFT/uniform) of the ground-state-to-ground-state $^{100}$Nb$\rightarrow$$^{100}$Mo $\beta$-decay spectrum. 
The DFT (uniform) finite-size correction was 
calculated using the values of $\la r\ra_{(2)}$ and $\la r\ra_{(2)}^r$ in Table~II that are labeled Mix (Uni). 
Both spectra are normalized to unity.
(Right) The FS correction $\delta_{FS}$ for the DFT and the (equal) uniform-density approximation.}
\end{figure}

In Fig.~(\ref{fig:FS-100Nb}) we compare the FS correction for the $^{100}$Nb$\rightarrow^{100}$Mo ground-state-to-ground-state transition using
the Zemach moments listed in Table~II with the result obtained using  Eqn.~(\ref{deltafs}) with $\la r \ra_{(2)}=\frac{36}{35}R$. 
The latter equation only holds true for equal charge and weak transition densities, and the value $\frac{36}{35}R$ for the Zemach moment
further requires the assumption of uniform densities.
The improved treatment of the densities leads to a small change in the FS correction, i.e., the slopes of the two lines in the right panel of Fig.~(\ref{fig:FS-100Nb}) differ by 12\%.
The left panel of Fig.~(\ref{fig:FS-100Nb}) shows that this leads to a maximum change in the shape of the
beta-decay spectrum of 0.5\%. 
The 12\% change in $\delta_{FS}$ translates into a 0.5\% change in the shape of the $\beta$-spectrum
because the FS correction comes in as a $(1+\delta_{FS}$) multiplicative spectrum correction 
 and so the ratio of the new to old normalized $\beta$-decay spectrum is a line of
slope $(\delta_{FS}^{new}-\delta_{FS}^{old})/2$ that goes through unity at $E_0/2$, where $E_0$ is the end-point energy.

\section{Summary}
\label{sec4}
The magnitudes of the expected reactor antineutrino spectra have recently been increased \cite{huber}, in part because of changes
in the treatment of the FS corrections to allowed nuclear $\beta$-decay.
In this work we have expressed the FS correction to allowed nuclear $\beta$-decay in terms of nuclear Zemach moments.
Using the
Skyrme-Hartree-Fock model we evaluated these Zemach moments for ground-state-to-ground-state transitions for a range of nuclei.
Our previous evaluations made the very simplified assumption of 
 uniform distributions for the normalized charge and weak transition densities determining these Zemach moments ~\cite{hayes1, friar1, friar2}.
 Despite the large differences between the shapes of the weak transition densities calculated using Hartree-Fock and those from model uniform densities,  the change in the Zemach moments was found to be relatively small, only $\sim$ 10-15\% for medium and heavy nuclei. 
  
Many of the allowed transitions contributing to 
reactor antineutrino spectra involve transitions
to excited states of the daughter nucleus. The present Hartree-Fock model is only designed 
to calculate ground-state wave functions, so we have not calculated such transitions.
However, for allowed $\beta$-decay transitions to excited states that are dominated by a weak transition density
involving neutron and proton orbitals of the same quantum numbers $ n \ell j$, or their spin-orbit partners, 
all of the conclusions drawn in the present work hold.  
However, this is not the case for the $\sim$30\% of the transitions making up fission antineutrino spectra that result from
forbidden $\beta$-decays, because the FS corrections considered in the present work only apply to allowed $\beta$-decays.

\begin{acknowledgments}

X.B. Wang wishes to thank the National Natural Science Foundation of China under Grant Nos. 11505056 and 11547312, and China Scholarship Council (201508330016) for supporting his research. A.C. Hayes thanks the Los Alamos National Laboratory LDRD program. 
\end{acknowledgments}
\section{Appendix: Zemach Densities and Moments}

The second form in Eqn.~(\ref{Z1}) defines $\la r \ra_{(2)} $ as the first radial moment of a convoluted or Zemach density
\begin{equation}
\rho_{(2)} (s) =  \int d^3 x\;  \rho\wlow (x)\,\rho\clow (\mid \vx -\vs \mid)\, . \label{rho2}
\end{equation}
The transformation from the first form to the second form  in Eqn.~(\ref{Z1}) is obtained by changing the inner variable $\vs$ to $\vr - \vs$ and then interchanging the order of integration.
The integral above can be simplified to a two-dimensional integral by first choosing the z-axis of the $\hat{s}$ coordinate system along the vector $\hat{x}$. This reduces the integration to two variables: $x$ and $\cos{\theta} = \hat{x} \cdot \hat{s}$. The integral over $\cos{\theta}$ can be simplified by changing variables from $\mu = \cos{\theta}$ to $\mid \vx - \vs \mid =\sqrt{x^2 +s^2 -2 x s \mu}$, which produces\cite{friar1}
\begin{equation}
\frac{1}{4 \pi} \int d\Omega_x \; \rho_ {\rm ch} (\mid \vx -\vs \mid)\, = \haf \int_{-1}^1 d\mu\; \rho_ {\rm ch} (\sqrt{x^2+s^2 -2xs \mu}) = \frac{1}{2 x s} \int_{\mid x-s \mid}^{x+s} dz\; z\; \rho\clow (z)\, , \label{ave}
\end{equation}
and finally
\begin{equation}
\rho_{(2)} (s) =  \frac{4 \pi}{2 s} \int_0^{\infty} dx\, x \;  \rho\wlow (x) \,\int_{\mid x-s \mid}^{x+s} dz\; z\; \rho\clow (z)\, . \label{rho2x}
\end{equation}
It follows immediately from Eqn.~(\ref{rho2})  that $ \int d^3s \; \rho_{(2)} (s)\,= 1$ if the individual densities are so normalized.

A similar treatment of Eqn.~(\ref{Zr}) allows it to be written as the first radial moment of a  Zemach-type density $ \rho_{(2)}^r (s)$\begin{equation}
\rho_{(2)}^r (s) =  \int d^3 x\; \, \rho\wlow (x)\; x\; \frac{\partial}{\partial x} \; \rho_ {\rm ch} (\mid \vx -\vs \mid)\, = \frac{4 \pi}{2 s} \int_0^{\infty} dx\, x^2 \;  \rho\wlow (x) \, x \frac{\partial}{\partial x} \frac{1}{x} \int_{\mid x-s \mid}^{x+s} dz\; z\; \rho_ {\rm ch} (z)\, . \label{rho2r}
\end{equation}
Performing the derivative leads to a relatively simple result
\begin{equation}
\rho_{(2)}^r (s)\,= -\rho_{(2)} (s) + \frac{1}{2s} \int d^3x \; \rho\wlow (x) \left[ (x+s) \rho_ {\rm ch} (x+s) -\;(x-s) \rho_ {\rm ch} (\mid x-s \mid) \right] \,  .\label{rho2rx}
\end{equation}
We note that $ \int d^3s \; \rho_{(2)}^r (s)\,= 0$, which follows from the first form in Eqn.~(\ref{rho2r}).

The relationship between the first Zemach moment and Holstein's form factor expression is
\begin{equation}
\la r \ra_{(2)} = -\frac{4}{\pi}\int_0^\infty \frac{dq}{q^2} (F\wlowF(q^2) F\clowF (q^2) - 1)\, , \label{FS}
\end{equation}
where $F\wlowF (q^2) = \frac{1}{(2 \pi)^3 } \int_0^{\infty} d^3 x \,\rho\wlow (x) e^{-i \vq \cdot \vx} $ and analogously for $F\clowF (q^2)$. We can verify this by inserting both form factor definitions into the $q$-integral. Noting that the integrand in Eqn.~(\ref{FS}) is always finite, we have
\begin{equation}
 -\frac{4}{\pi}\int_0^\infty \frac{dq}{q^2} (F\wlowF(q^2) F\clowF (q^2) - 1) = -\frac{1}{\pi^2} \int \frac{d^3 q}{q^4} (F\wlowF (q^2) F\clowF (q^2) - 1) = \int d^3 x\, \rho\clow (x) \int d^3 y\, \rho\wlow (y) I (z) \, , \label{FSX}
\end{equation}
where $\vz = \vx - \vy$ and
\begin{equation}
I(z) = -\frac{1}{\pi^2} \int \frac{d^3 q}{q^4} (e^{i \vq \cdot \vz} - 1) = -\frac{4}{\pi z} \int_0^{\infty} \frac{dq}{q^3} (\sin{(qz)} - qz) =  -\frac{4 z}{\pi} \int_0^{\infty} \frac{dt}{t^3} \left (\sin{(t)} - t \right)\, .
\end{equation}
We twice integrate-by-parts the coefficient of the bracketed term in the final integral, which produces $-\frac{\pi}{4}$ for the integral and thus $I(z) = z$. Inspection of the last form of Eqn.~(\ref{FSX}) demonstrates that the result is indeed $\la r \ra_{(2)} $.

The form-factor form of $\la r \ra_{(2)}^r $ is given by
\begin{equation}
\la r \ra_{(2)}^r = - \frac{8}{\pi} \int_0^{\infty} dq\; F\clowF (q^2) \frac{d\,  F\wlowF (q^2)}{d q^2 \hphantom{CCC}}\, , \label{r2r}
\end{equation}
which can be verified in much the same way that Eqn.~(\ref{FS}) was verified. We need the following trick for rewriting the derivative
\begin{equation}
 -\frac{8}{\pi}\int_0^\infty dq\, F\clowF (q^2) \frac{d\,  F\wlowF (q^2)}{\!d q^2 \hphantom{CCC}} = -\frac{1}{\pi^2} \int \frac{d^3 q}{q^4}  F\clowF (q^2)\, \vq \cdot \vnabla_q F\wlowF (q^2)\,.
\end{equation}
The factor of $\vq \cdot \vnabla_q$ acting on the Fourier exponential in $F\wlowF (q^2)$ (defined below Eqn.~(\ref{FS})) produces $ \vy \cdot \vnabla_y$. The rest of the derivation is straightforward.

A uniform model credibly approximates the charge density, but not the weak transition density.
Reference~\cite{bible} treats the weak transition density as a form factor, which they expand as a power series in $q^2$ (see Eqn.~(6.74)).  If only the spatial extent of the weak density is significant, we can keep only the leading-order structure term (viz., their $n=1$ term) in the form $F\wlowF (q^2) \cong 1-q^2 \la r^2 \ra\wlow /6$, where $\la r^2 \ra\wlow$ is the mean-square radius of the weak density. Inserting this approximation into Eqn.~(\ref{FS}) leads to
\begin{equation}
\la r \ra_{(2)}^{BB} \cong -\frac{4}{\pi}\int_0^\infty \frac{dq}{q^2} \left(\left(1-\frac{q^2}{6} \la r^2 \ra\wlow \right)\, F\clowF (q^2) - 1\right) = -\frac{4}{\pi}\int_0^\infty \frac{dq}{q^2} (F\clowF (q^2) - 1) +\frac{2 \la r^2 \ra\wlow}{3 \pi}\int_0^\infty d q\,  F\clowF (q^2)  \, , \label{FSA}
\end{equation}
where the superscript ``$BB$" on the moment refers to the authors of Ref.~\cite{bible}.
The first term is just Eqn.~(\ref{FS}) with $F\wlowF = 1$ and equals $\la r \ra\clow$. Inserting the definition of $F\clowF$ into the second term and performing the $q$-integral produces $\frac{\pi}{2} \la \frac{1}{r} \ra\clow$ for that integral. Our final result is
\begin{equation}
\la r \ra_{(2)}^{BB} \cong \la r \ra\clow + \frac{\la r^2 \ra\wlow  \la \frac{1}{r} \ra\clow}{3} \, . \label{FSBB}
\end{equation}
We can make the same approximation for $\la r \ra_{(2)}^r$ in Eqn.~(\ref{r2r}), which produces
\begin{equation}
\la r \ra_{(2)}^{r-BB} \cong  \frac{2 \la r^2 \ra\wlow  \la \frac{1}{r} \ra\clow}{3} \, . \label{FSXX}
\end{equation}

How well these approximations work can be most easily investigated using two identical distributions. For the uniform distribution we have $\la r \ra = 3R/4$, $\la r^2 \ra = 3R^2/5$, $\la 1/r \ra =3/(2R)$ and $\la r \ra_{(2)} = 36 R/35$. Equation~(\ref{FSBB}) then produces $\la r \ra_{(2)}^{BB} = R (3/4 +3/10)$ and $ \la r \ra_{(2)}^{BB}/\la r \ra_{(2)} =49/48$. For this case the approximation is about 2\% too large. The second term in Eqn.~(\ref{FSBB}) increases the Zemach moment 40\%, which is about the difference in the sizes of $\la r  \ra_{(2)}^{\rm Ch}$ and $\la r \ra^{\rm Ch}$ in Tables~I and~II. This also illustrates the substantial effect of how a second density increases radial moments by smearing out the distribution, and how changing the extent of that second density changes $\la r \ra_{(2)}$.

Using the moments calculated in Table~V of Ref.~\cite{friar1} the very different exponential distribution produces  $ \la r \ra_{(2)}^{BB}/\la r \ra_{(2)} =8/7$, which is about 14\% too large, while the Gaussian distribution produces results about 6\% too large.

\bibliographystyle{unsrt}

\end{document}